\documentclass[11pt,onecolumn]{IEEEtran}
\usepackage{epsfig}
\usepackage{subfigure}
\usepackage{times}
\usepackage{color}

\newcommand{\beq}{\begin{equation}}
\newcommand{\eeq}{\end{equation}}

\begin{document}     
\voffset 1cm

\title {\Large{\bf Capacity Achieving Code Constructions for Two Classes of ($d,k$) Constraints$^{\dag}$\footnote{$^{\dag}$This work was supported by Seagate Research.}}}
\large\author{\emph{Yogesh Sankarasubramaniam\hspace{0.5cm}Steven W. McLaughlin}\\                          
\normalsize School of Electrical \& Computer Engineering\\
\normalsize Georgia Institute of Technology, Atlanta, GA 30332\\
\normalsize E-mail:\{yogi,swm\}@ece.gatech.edu
}

\date{}

\maketitle 

\vspace{-0.2in}

\begin{abstract}
In this paper, we present two low complexity algorithms that achieve
capacity for the noiseless ($d,k$) constrained channel when $k=2d+1$,
or when $k-d+1$ is not prime. The first algorithm, \emph{symbol sliding}, is a generalized version of the bit flipping algorithm introduced by Aviran \emph{et al} \cite{Aviran04}. In addition to achieving capacity for ($d,2d+1$) constraints, it comes close to capacity in other cases. The second algorithm
is based on interleaving, and is a generalized version of the bit stuffing algorithm  
introduced by Bender and Wolf \cite{Bender93}. This method uses fewer than $k-d$ biased bit streams to achieve capacity for ($d,k$) constraints with $k-d+1$ not prime. In particular, the encoder for ($d,d+2^m-1$) constraints, $1\le m<\infty$, requires only $m$ biased bit streams.
\end{abstract}


\section{Introduction}

A binary sequence is said to be ($d,k$) constrained if successive ones
are separated by at least $d$ and at most $k$ consecutive zeros.
There is a long history of
the use of ($d,k$) constrained codes and they are part of virtually all
magnetic and optical disk recording systems today. The $d$ constraint is used to regulate intersymbol interference and the $k$ constraint is important for timing recovery. Over the years, gains in storage density, manufacturing tolerances and system margins have been possible with the use of ($d,k$) codes (see \cite{Immink} for an overview).

The basic issues in coding for a constrained channel are \emph{rate} and
\emph{complexity}. With the assumption of a noiseless ($d,k$) constrained channel, the \emph{Shannon capacity}, $C(d,k)$, is given by \cite{Shannon}
\begin{displaymath}
C(d,k)=\log_2\lambda,
\end{displaymath}
\noindent where $\lambda$ is the positive, real root\footnote{Sometimes the notation $\lambda_{d,k}$ is used for emphasis when the constraint ($d,k$) is not already clear from context.} of the
characteristic equation $H_{d,k}(z)=1$, and $H_{d,k}(z)$ is the characteristic polynomial of the ($d,k$) constraint, given by
\beq
\label{charpoly}
H_{d,k}(z)=\left\{
\begin{array}{ll}
\sum_{j=d+1}^{k+1}z^{-j} & {\rm when~} k<\infty\\ 
z^{-1}+z^{-(d+1)} & {\rm when~} k=\infty
\end{array}
\right.
\eeq

\noindent $C(d,k)$ is an upper bound on the information rate, $R(d,k)$, of any encoding algorithm. The encoder efficiency $E(d,k)=R(d,k)/C(d,k)$ measures how close the code is to capacity. Clearly, the challenge is to design low complexity codes with high efficiency. Of particular interest are \emph{optimal} codes/algorithms that are $100\%$ efficient. 

Our aim in this work is to improve upon techniques that use very simple encoding ideas. In this regard, Bender and Wolf \cite{Bender93} first proposed the \emph{bit stuffing}
algorithm to generate ($d,k$) constrained sequences. They showed that controlled insertion of bits into an appropriately biased, independent and identically distributed ($i.i.d$) bit stream, is asymptotically optimal for the ($d,\infty$) and ($d,d+1$) constraints and near-optimal for other constraints. More recently, the \emph{bit flipping} algorithm \cite{Aviran04} was shown to improve bit stuffing rates for most ($d,k$) constraints and additionally achieve ($2,4$) capacity. For all values of ($d,k$), $k\ne d+1$, $k\ne \infty$ and ($d,k$)$\ne$($2,4$), bit flipping was shown to be suboptimal. 

As a first step, both the bit stuffing and bit flipping algorithms use a \emph{distribution transformer} (DT) to introduce a bias into the unbiased ($Pr\{0\}=0.5$) binary, $i.i.d$ input stream. This has the effect of better conforming the input to the constraint before the actual bit insertion is performed. Wolf \cite{Aviran04} observed that with the use of multiple such DTs, one could, in theory, generate enough degrees of freedom to produce optimal ($d,k$) sequences for all $0\le d<k$. More precisely, $k-d$ DTs were shown to be sufficient for any given ($d,k$) constraint, $k<\infty$.

In this work, we introduce two code constructions that improve upon the aforementioned encoding algorithms. Our first construction is the \emph{symbol sliding} algorithm which improves bit stuffing and bit flipping rates while still using only a single DT. We prove the optimality of the proposed algorithm for all ($d,k$) constraints with $k=2d+1$, and show that bit stuffing and bit flipping can be derived as special cases of symbol sliding. Our second construction is based on \emph{interleaving} and uses fewer than $k-d$ DTs to achieve capacity for all ($d,k$) constraints with $k-d+1$ not prime. In the particular case of ($d,d+2^m-1$) constraints, our construction requires only $m=log_2(k-d+1)$ DTs.

The remainder of this paper is organized as follows. We begin by reviewing the bit stuffing and bit flipping algorithms in Section \ref{bitstu}. We provide an interpretation of matching phrase probabilities to those of the maxentropic sequence and motivate symbol sliding using the example of the ($1,3$) constraint. Next, in Section \ref{symsli}, we study the symbol sliding algorithm and prove its optimality for ($d,2d+1$) constraints. We then proceed to discuss code constructions using interleaving in Section \ref{interl} and conclude in Section \ref{conclu}.

\section{Background: Bit Stuffing and Bit Flipping}
\label{bitstu}

Both our code constructions are inspired by the simple concept of stuffing bits to satisfy ($d,k$) constraints. In order to gain the necessary understanding and motivation behind our proposed constructions, we first review the bit stuffing algorithm.

\subsection{The Bit Stuffing Algorithm}
\label{bstuf}

\begin{figure}[htb]
\centerline{
\begin{minipage}[h]{3in}
\centerline
{\psfig{figure=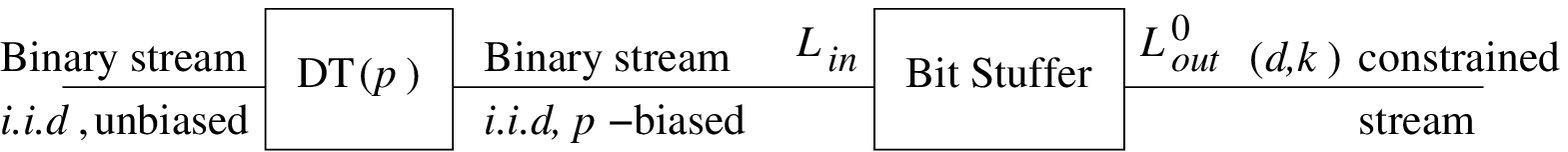,width=3in}}
\caption{\textbf{Block diagram of the bit stuffing encoder. {\rm DT($p$)} denotes a distribution transformer with bias $p$. {\rm $L_{in}$} denotes the average message word length at the input to the bit stuffer, and {\rm $L_{out}^o$} denotes the average output word length.}}
\label{bst}
\end{minipage}
\hspace*{1cm}
\begin{minipage}[h]{3in}
\centerline
{\psfig{figure=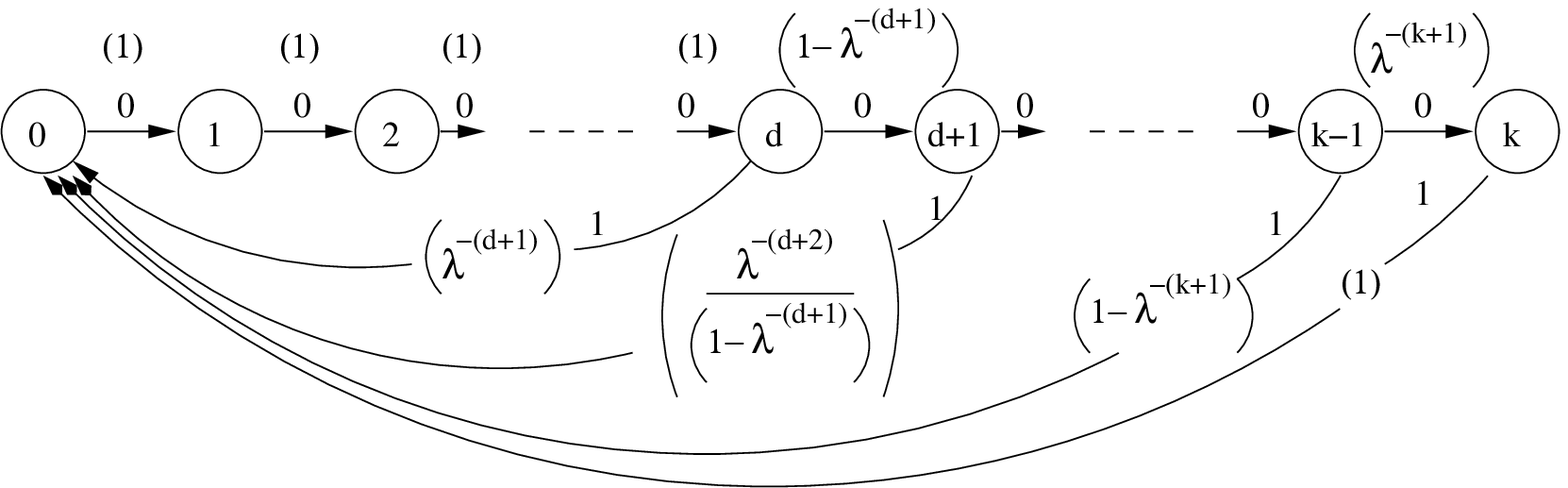,width=3in}}
\caption{\textbf{FSTD with maxentropic state transition probabilities in parentheses. The labels on directed edges indicate the output bit.}}
\label{maxentr}
\end{minipage}
}
\end{figure}

Bit stuffing \cite{Bender93} is a simple, but surprisingly efficient, algorithm to generate ($d,k$) sequences. The block diagram of the bit stuffing encoder is shown in Fig. \ref{bst}. It consists of a distribution transformer (DT) followed by a bit stuffer. The DT converts the unbiased ($Pr\{0\}=\frac{1}{2}$), binary, $i.i.d$ input stream into a $p$-biased (Pr$\left\{0\right\}=p$), binary, $i.i.d$ stream. This conversion occurs at an asymptotic rate penalty of $h(p)$ information bits, where $h(.)$ is the binary entropy function. However, with a suitable choice of $p$, the biasing can actually improve overall rates by better fitting input data to the constraint. 

The $p$-biased stream generated by the DT is then fed into the bit stuffer, which sequentially performs the following two operations
\begin{enumerate}
\item Insert a one after every run of $k-d$ consecutive zeros (skip this step if $k=\infty$)
\item Stuff $d$ zeros after every one
\end{enumerate}

\noindent The first operation produces a ($0,k-d$) constrained sequence, which then acts as input for the second operation. Stuffing $d$ zeros in the second operation translates the ($0,k-d$) constraint to the required ($d,k$) constraint. Both these operations are invertible. Hence, with a one-to-one implementation of the DT (see \cite{Jones} for a possible method), the bit stuffing decoder is a simple inverse of the encoder.

Bender and Wolf \cite{Bender93} showed that with a proper choice of bias $p$, the maximum average rate of the bit stuffing algorithm equals ($d,k$) capacity for $k=d+1$ and $k=\infty$, and is strictly less than capacity for all other cases. We now provide an alternate interpretation of their results. This is based on matching phrase probabilities and will help motivate the need for our proposed algorithm in Section \ref{symsli}. 

Consider the finite state transition diagram (FSTD) of a ($d,k$) constraint, as shown in Fig. \ref{maxentr}. Walks on the FSTD can be used to generate all possible ($d,k$) sequences. It is well known that there is a \emph{maxentropic} walk, where edges must be traversed according to a set of optimal state transitions in order to achieve the highest possible rate. A code achieves capacity if and only if it produces a walk on the FSTD with the maxentropic state transition probabilities (shown in parentheses in Fig. \ref{maxentr}). 

Alternatively, one can describe a ($d,k$) sequence by the concatenation of independent phrases from the finite set $\mathcal{X}=\left\{0^k1,0^{k-1}1,\ldots,0^{d-1}1,0^d1\right\}$, where $0^t1$ represents a sequence of $t$ zeros followed by a one. Each phrase corresponds to a cycle on the FSTD (see Fig. \ref{maxentr}) that begins and ends in State $0$. Note if $k=\infty$, then $\mathcal{X}=\left\{0,0^d1\right\}$ and the FSTD in Fig. \ref{maxentr} can be redrawn with exactly $d+1$ states. A code achieves capacity if and only if it generates constrained phrases with maxentropic probabilities. We can hence form a maxentropic phrase probability vector, ${\bf \Lambda}$, which is given by \cite{Zehavi}
\beq
{\bf \Lambda}=\left[\lambda^{-(k+1)}~ \lambda^{-(k)} \ldots \lambda^{-(d+2)}~ \lambda^{-(d+1)}\right],
\eeq
\noindent where $\lambda^{-t}$ denotes the maxentropic probability of occurrence of a ($d,k$) constrained phrase of length $t$, namely $0^{t-1}1$. With the bit stuffing algorithm, we can form the corresponding phrase probability vector 
\beq
\textbf{v}^0=\left[v^0_0~ v^0_1 \ldots v^0_{k-d-1}~ v^0_{k-d}\right],
\eeq
\noindent where $v^0_{i}$ denotes the probability of occurrence of the phrase $0^{k-i}1$. Table \ref{bitstuff} lists the output ($d,k$) constrained phrases and corresponding message words at the input to the bit stuffer (see Fig. \ref{bst}). Recall that the bit stuffer input is $p$-biased, thereby yielding the corresponding phrase probabilities $v^0_i$.
\begin{table*}[ht!]
\caption{Bit Stuffing Phrase Probabilities}
\label{bitstuff}
\begin{center} 
\small 
\begin{tabular}{|c|c|c|c|} \hline
{\bf Index} & {\bf ($d,k$) constrained} & {\bf Corresponding} & {\bf Phrase probability}\\
($i$) & {\bf phrase} & {\bf message word} & ($v^0_i$)\\ \hline \hline
$0$ & $0^k1$ & $0^{(k-d)}$ & $p^{(k-d)}$ \\ \hline
$1$ & $0^{(k-1)}1$ & $0^{(k-d-1)}1$ & $p^{(k-d-1)}(1-p)$ \\ \hline
\vdots & \vdots & \vdots & \vdots\\ \hline
t&$0^{(k-t)}1$ & $0^{(k-d-t)}1$ & $p^{(k-d-t)}(1-p)$ \\ \hline
\vdots & \vdots & \vdots & \vdots \\ \hline
$k-d-1$ & $0^{(d+1)}1$ & $01$ & $p(1-p)$ \\ \hline
$k-d$ & $0^d1$ & $1$ & $1-p$ \\ \hline
\end{tabular}
\end{center}
\end{table*}

Hence, the bit stuffing algorithm achieves capacity if and only if $\textbf{v}^0={\bf \Lambda}$. It can be verified that for ($d,d+1$) and ($d,\infty$) constraints, $\textbf{v}^0$ exactly matches ${\bf \Lambda}$ with $p=\lambda_{d,d+1}^{-(d+2)}$ and $p=\lambda_{d,\infty}^{-1}$, respectively. The following proposition restates a result of Bender and Wolf \cite{Bender93}.

\smallskip
\noindent \emph{Proposition 1:} For $d\ge 0$, $d+2\le k<\infty$, $\textbf{v}^0\ne{\bf \Lambda}$

\smallskip
\noindent Proposition 1 implies that the maximum average bit stuffing rate is strictly less than capacity for $d+2\le k<\infty$. Our objective now is to improve bit stuffing rates for $d+2\le k<\infty$ while maintaining similar implementation complexity. We will show that this can be done by switching the bit stuffing phrase probabilities to better match the maxentropic vector ${\bf \Lambda}$. As a first step, we show how this idea leads to the bit flipping algorithm and then generalize to symbol sliding in Section \ref{symsli}.

\subsection{The Bit Flipping Algorithm}
\label{bitfli}

Consider a DT bias of $p$ greater than $0.5$. This means that a $0$ is more likely than a $1$ at the input to the bit stuffer (see Fig. \ref{bst}). Recall that our goal is to match the phrase probability vector, $\textbf{v}^0$, to the maxentropic vector ${\bf \Lambda}$. Looking at indices $i=0$ and $i=1$ in Table \ref{bitstuff}, we note that $v^0_0=p^{(k-d)}>v^0_1=p^{(k-d-1)}(1-p)$, but the corresponding maxentropic probabilities are related as $\lambda^{-(k+1)}<\lambda^{-(k)}$. This suggests that switching the roles of $v^0_0$ and $v^0_1$ should result in a better match with ${\bf \Lambda}$, thereby improving bit stuffing rates. Hence, we would like to replace the bit stuffer in Fig. \ref{bst} by a constrained encoder that sequentially performs the following three operations on the biased bit stream
\begin{enumerate}
\item Track the run-length ($\rho$) of consecutive zeros, including the current bit (skip this step if $k=\infty$)
\begin{itemize}
\item If current bit is zero and $\rho=k-d-1$, flip the next bit, reset $\rho$ and goto 1)
\item If current bit is one and $\rho<k-d-1$, reset $\rho$ and goto 1)
\end{itemize}
\item Insert a one after every run of $k-d$ consecutive zeros (skip this step if $k=\infty$)
\item Stuff $d$ zeros after every one
\end{enumerate}

\noindent The first operation performs the bit flipping (change ones to zeros and \emph{vice versa}), which switches the roles of $v^0_0$ and $v^0_1$. The second and third operations are identical to the bit stuffer operations described in Section \ref{bstuf}. Since the bit flipping operation is invertible, the decoder once again is simply the encoder components arranged in the reverse order. 

The algorithm described above is precisely the bit flipping algorithm proposed by Aviran \emph{et al} \cite{Aviran04}. Their main results are summarized in the following two propositions

\smallskip
\noindent \emph{Proposition 2:} For $d\ge 1$, $d+2\le k<\infty$, the bit flipping algorithm achieves greater maximum average rate than the bit stuffing algorithm.

\smallskip
\noindent \emph{Proposition 3:} For $d\ge 0$, $d+2\le k<\infty$, the bit flipping algorithm is optimal if and only if $d=2$ and $k=4$.
\smallskip

\noindent Proposition 2 mainly depends on the following two facts
\begin{itemize}
\item The average bit flipping rate is greater than the average bit stuffing rate for all values of bias $p$ greater than $0.5$
\item The rate maximizing bit stuffing bias is greater than $0.5$ for all $d\ge 2$, $d+2\le k<\infty$
\end{itemize}
\noindent Proposition 3 states that the new phrase vector, say $\textbf{v}^1$, formed by swapping the roles of $v^0_0$ and $v^0_1$ in $\textbf{v}^0$, exactly matches ${\bf \Lambda}$ only for the ($2,4$) constraint. As will be seen later, this optimality of the bit flipping algorithm is possible only because of the binary capacity equality $C(2,4)=C(1,2)$. 

\subsection{Motivating Example: The ($1,3$) Constraint}
\label{motexa}

Thus far, we have seen a phrase probability interpretation of bit stuffing and how switching two entries of the phrase probability vector $\textbf{v}^0$ improved rates with the bit flipping algorithm. This prompts us to generalize the idea of switching phrase probabilities to better match the maxentropic vector ${\bf \Lambda}$. The following example of the ($1,3$) constraint motivates this idea. 
\begin{table*}[ht!]
\caption{Phrase probabilities for the (1,3) constraint}
\label{ex13}
\begin{center} 
\small 
\begin{tabular}{|c|c|c|c|c|c|} \hline
{\bf Index ($i$)} & {\bf (1,3) constrained} & {\bf Maxentropic} & {\bf Bit stuffing} & {\bf Bit flipping} & {\bf Symbol sliding with index $2$}\\
& {\bf phrase} & {\bf prob. (${\bf \Lambda}(i)$)} & ($v^0_i$) & ($v^1_i$) & ($v^2_i$)\\ \hline \hline
0 & $0^31$ & $\lambda^{-4}$ & $p^{2}$ & $p(1-p)$ & $p(1-p)$\\ \hline
1 & $0^21$ & $\lambda^{-3}$ & $p(1-p)$ & $p^2$ & $1-p$\\ \hline
2 & $0^11$ & $\lambda^{-2}$ & $1-p$ & $1-p$ & $p^2$\\ \hline
\end{tabular}
\end{center}
\end{table*}  

Consider the phrase probabilities listed in Table \ref{ex13}. From Proposition 1, it follows that the maximum average bit stuffing rate is strictly less than ($1,3$) capacity. Proposition 3 states that ($1,3$) bit flipping rates are also suboptimal. Now consider the phrase probabilities $v_i^2$ as listed in the last column of Table \ref{ex13}. We call this \emph{symbol sliding} with index $2$. This means that the role of $v^0_0$ is slid down to that of $v^0_2$ (index 2) with $v^0_2$ and $v^0_1$ being pushed up an index each, thus yielding the phrase probability vector $\textbf{v}^2=[v^2_0~ v^2_1~ v^2_2]$. It can be seen that with a bias of $p=\lambda^{-1}$, $\textbf{v}^2$ exactly matches ${\bf \Lambda}$, and the average rate is equal to the ($1,3$) capacity. Hence, symbol sliding with index $2$ achieves capacity for the ($1,3$) constraint where both bit stuffing and bit flipping fall short. This prompted us to study symbol sliding in greater depth.

\section{Construction 1: The Symbol Sliding Algorithm}
\label{symsli}

The main idea behind symbol sliding is to switch the roles of bit stuffing phrase probabilities so as to better match the phrase probability vector to the maxentropic vector ${\bf \Lambda}$. Symbol sliding is hence a function of a sliding index, $j, 0\le j\le k-d$, for a given ($d,k$) constraint. Symbol sliding with index $j$ involves sliding down $v_0^0$ from index $i=0$ to $i=j$ and moving each of $v_1^0,v_2^0,\dots,v_j^0$ up an index each, to yield the phrase probability vector $\textbf{v}^j=[v_0^j~ v_1^j~ \ldots v_{k-d}^j]$. Table \ref{list} provides the full list of bit stuffing, bit flipping, symbol sliding and maxentropic phrase probabilities.

\begin{figure}[ht!]
  \centerline{\psfig{figure=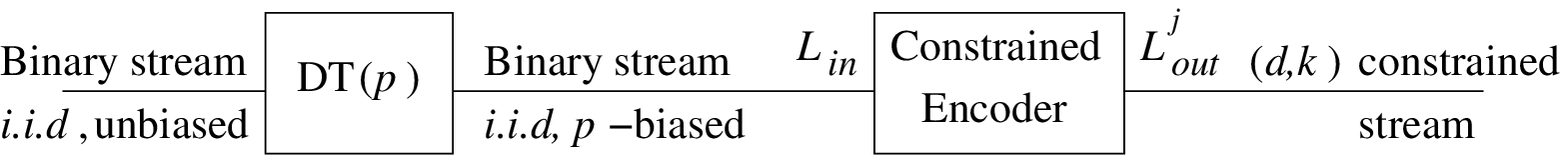,width=4in}}
  \caption{\textbf{Block diagram of the symbol sliding encoder. {\rm $L_{in}$} denotes the average message word length at the input to the constrained encoder. {\rm $L_{out}^j$} denotes the average output word length for sliding index $j$.}}
  \label{cenc}
\end{figure}

The symbol sliding encoder is shown in Fig. \ref{cenc}. It has a similar set up to the bit stuffing encoder with the bit stuffer being replaced by a \emph{constrained encoder} that sequentially performs the following two operations on the biased bit stream
\begin{enumerate}
\item Track the run-length ($\rho$) of consecutive zeros, including the current bit (skip this step if $k=\infty$)
\begin{itemize}
\item If current bit is zero and $\rho=k-d$, replace the run of $k-d$ zeros with the phrase $0^{k-d-j}1$, reset $\rho$ and goto 1)
\item If current bit is one and $k-d-j\le \rho\le k-d-1$, insert a zero before the current bit, reset $\rho$ and goto 1)
\item If current bit is one and $\rho<k-d-j$, reset $\rho$ and goto 1)
\end{itemize}
\item Stuff $d$ zeros after every one
\end{enumerate}
\noindent The first operation produces a ($0,k-d$) constrained sequence with the appropriate phrase matching and the second operation translates this to a ($d,k$) constraint by stuffing $d$ zeros. The latter is identical to the corresponding bit stuffing operation. 

The constrained decoder is a simple inverse of the constrained encoder. It sequentially performs the following operations on the ($d,k$) sequence
\begin{enumerate}
\item Remove the $d$ stuffed zeros after every one
\item Track the run-length ($\rho$) of consecutive zeros, including the current bit (skip this step if $k=\infty$)
\begin{itemize}
\item If current bit is one and $k-d-j+1\le \rho\le k-d$, remove the stuffed zero before the current bit, reset $\rho$ and goto 2)
\item If current bit is one and $\rho=k-d-j$, replace the phrase $0^{k-d-j}1$ by a run of $k-d$ zeros, reset $\rho$ and goto 2)
\item If current bit is one and $\rho<k-d-j$, reset $\rho$ and goto 2)
\end{itemize}
\end{enumerate}

Let us denote by SS($j$), the symbol sliding algorithm with index $j$. It can be seen from Table \ref{list} that SS(0) and SS(1) are identical to the bit stuffing and bit flipping algorithms, respectively. Hence, bit stuffing and bit flipping are two special cases of the symbol sliding algorithm. We now summarize some important properties and prove the optimality of symbol sliding for certain constraints. 

\subsection{Properties of Symbol Sliding}
\label{proper}

\medskip
\newtheorem{lemma}{Lemma}
\begin{lemma}
\label{letdol}
Let $0\le d<k<\infty$. Then, the maximum average rate achieved by SS($j$) equals ($d,k$) capacity when $k=2d+1$ and sliding index $j=k-d=d+1$.
\end{lemma}

\smallskip

\begin{proof}
We will show that SS($j$) generates maxentropic ($d,k$) constrained phrases when $k=2d+1$ and $j=d+1$. We start with a result of Ashley and Siegel \cite{Ashley}, which states that the capacity of the ($d,2d+1$) constraint is identical to that of the ($d+1,\infty$) constraint. Hence $\lambda_{d,2d+1}$ is the positive, real root of each of the following two characteristic equations
\begin{eqnarray}
\label{twoinone}
\sum_{l=d+1}^{2d+2} z^{-l}=1 \nonumber\\
z^{-1}+z^{-(d+2)}=1
\end{eqnarray}

\noindent Now, let the sliding index $j=k-d=d+1$. Consider a bias $p=\lambda_{d,2d+1}^{-1}$. Then, we have 
\begin{eqnarray}
& & v^{d+1}_{k-d}=p^{k-d}=p^{d+1}=\lambda_{d,2d+1}^{-(d+1)}\\
\label{sec}
& & v^{d+1}_{k-d-1}=1-p=1-\lambda_{d,2d+1}^{-1}=\lambda_{d,2d+1}^{-(d+2)}\\
& & v^{d+1}_{k-d-i}=p^{i-1}(1-p)=\lambda_{d,2d+1}^{-(d+i+1)}, \quad 2\le i\le k-d
\end{eqnarray}

\noindent where (\ref{sec}) follows from (\ref{twoinone}). Hence, we have $v^{d+1}_{i}=\lambda_{d,2d+1}^{-(k-i+1)}$, for all $0\le i\le k-d$, whereby $\textbf{v}^{d+1}={\bf \Lambda}$. This proves the lemma. 
\end{proof}

\medskip

\newtheorem{theorem}{Theorem}
\begin{theorem}
\label{forzer}
For $0\le d<k$, the maximum average rate achieved by SS($j$) equals the ($d,k$) capacity only in the following cases
\begin{enumerate}
\item $j=0$, $k=d+1$
\item $j=1$, $k=d+1$
\item $j=1$, $d=2,k=4$
\item $j=k-d$, $k=2d+1$
\item $j\ge 0$, $k=\infty$
\end{enumerate}
\noindent For all other values of ($d,k$), the maximum average rate of SS($j$) is strictly less than capacity for each $j, 0\le j\le k-d$.
\end{theorem} 

\smallskip

\begin{proof}
We wish to find constraints ($d,k$) for which $\textbf{v}^{j}={\bf \Lambda}$ for some $0\le j\le k-d$. We first note that when there is no $k$ constraint, $i.e.$, $k=\infty$, then the symbol sliding operations reduce to simply stuffing $d$ zeros after every one in the biased bit stream. This is identical to the corresponding bit stuffing operation, which has been shown to achieve capacity for ($d,\infty$) constraints \cite{Bender93}. Case 5) in the theorem statement now follows. In the remainder of this proof, we focus only on ($d,k$) constraints with $k<\infty$. 

Depending on the value of $j$, we have the following four cases.

\smallskip

\noindent {\bf Case 1:} $j=0$

This is identical to the bit stuffing algorithm. Let us first consider $k>d+1$. For any such given ($d,k$) constraint, the following must hold (see Table \ref{list}) in order for $\textbf{v}^{0}={\bf \Lambda}$. 
\begin{eqnarray}
\label{un}& & p=\lambda^{-1}\\
\label{deux}& & 1-p=\lambda^{-(d+1)}\\
\label{trois}& & p^{k-d}=\lambda^{-(k+1)}
\end{eqnarray}

\noindent (\ref{un}) and (\ref{deux}) together imply that $\lambda^{-1}+\lambda^{-(d+1)}=1$. However, this means that $\lambda$ is a root of the characteristic ($d,\infty$) equation, $H_{d,\infty}=1$. Hence,  (\ref{un}) and (\ref{deux}) cannot be simultaneously satisfied for any finite $k>d+1$. This leads us to Proposition 1 which was stated without proof in Section \ref{bitstu}.
 
Next, we look at $k=d+1$. In this case, we only have two possible phrases corresponding to indices $i=0,1$ in Table \ref{list}. It can be seen that a bias of $p=\lambda^{-(d+2)}$ is optimal. This yields Case 1) of the theorem statement.

\begin{table*}[ht!]
\caption{Maxentropic phrase probabilities along with those of the bit stuffing, bit flipping and symbol sliding algorithms}
\label{list}
\begin{center} 
\small 
\begin{tabular}{|c|c|c|c|c|c|} \hline
{\bf Index ($i$)} & {\bf ($d,k$) constrained} & {\bf Maxentropic} & {\bf Bit stuffing} & {\bf Bit flipping} & {\bf Symbol sliding with index $j$}\\
& {\bf phrase} & {\bf prob. (${\bf \Lambda}_{d,k}(i)$)} & ($v^0_i$) & ($v^1_i$) & ($v^j_i$)\\ \hline \hline
0 & $0^k1$ & $\lambda_{d,k}^{-(k+1)}$ & $p^{(k-d)}$ & $p^{(k-d-1)}(1-p)$ & $p^{(k-d-1)}(1-p)$\\ \hline
1 & $0^{(k-1)}1$ & $\lambda_{d,k}^{-(k)}$ & $p^{(k-d-1)}(1-p)$ & $p^{(k-d)}$ & $p^{(k-d-2)}(1-p)$\\ \hline
\vdots & \vdots & \vdots & \vdots & \vdots & \vdots\\ \hline
$j-1$ & $0^{(k-j+1)}1$ & $\lambda_{d,k}^{-(k-j+2)}$ & $p^{(k-d-j+1)}(1-p)$ & $p^{(k-d-j+1)}(1-p)$ & $p^{(k-d-j)}(1-p)$\\ \hline
$j$ & $0^{(k-j)}1$ & $\lambda_{d,k}^{-(k-j+1)}$ & $p^{(k-d-j)}(1-p)$ & $p^{(k-d-j)}(1-p)$ & $p^{(k-d)}$\\ \hline
$j+1$ & $0^{(k-j-1)}1$ & $\lambda_{d,k}^{-(k-j)}$ & $p^{(k-d-j-1)}(1-p)$ & $p^{(k-d-j-1)}(1-p)$ & $p^{(k-d-j-1)}(1-p)$\\ \hline
\vdots & \vdots & \vdots & \vdots & \vdots & \vdots\\ \hline
$k-d-1$ & $0^{(d+1)}1$ & $\lambda_{d,k}^{-(d+2)}$ & $p(1-p)$ & $p(1-p)$ & $p(1-p)$\\ \hline
$k-d$ & $0^d1$ & $\lambda_{d,k}^{-(d+1)}$ & $1-p$ & $1-p$ & $1-p$\\ \hline
\end{tabular}
\end{center}
\end{table*}

\smallskip

\noindent {\bf Case 2:} $j=1$

This is identical to the bit flipping algorithm. We first consider $k>d+2$. For any such given ($d,k$) constraint, the following must hold (see Table \ref{list}) in order for $\textbf{v}^{1}={\bf \Lambda}$.
\begin{eqnarray}
\label{un2}& & p=\lambda^{-1}\\
\label{deux2}& & 1-p=\lambda^{-(d+1)}\\
\label{trois2}& & p^{k-d}=\lambda^{-k}\\
\label{quatre2}& & p^{k-d-1}(1-p)=\lambda^{-(k+1)}
\end{eqnarray}

\noindent (\ref{un2}) and (\ref{deux2}) together imply that $\lambda^{-1}+\lambda^{-(d+1)}=1$. As in the previous case, this is impossible unless $k=\infty$. 

Next, let $k=d+2$. As before, from Table \ref{list}, we obtain the following conditions in order for $\textbf{v}^{1}={\bf \Lambda}$.
\begin{eqnarray}
\label{deux23}& & 1-p=\lambda^{-(d+1)}\\
\label{trois23}& & p^{2}=\lambda^{-(d+2)}\\
\label{quatre23}& & p(1-p)=\lambda^{-(d+3)}
\end{eqnarray}

\noindent From (\ref{trois23}) we have $p=\lambda^{-(\frac{d}{2}+1)}$. Using this and (\ref{deux23}) in (\ref{quatre23}), we see that $\frac{d}{2}+1+d+1=d+3$ or $d=2$. This implies that SS(1) is optimal for the ($2,4$) constraint, as stated in Case 3) of the theorem. 

Finally, let $k=d+1$. This means that we only have indices $i=0,1$ in Table \ref{list}. The bit flipping algorithm in this case is exactly the bit stuffing algorithm run on the corresponding flipped (ones changed to zeros and \emph{vice versa}) biased bit stream. Hence, for any bit stuffing bias $p$, a bit flipping bias of ($1-p$) achieves the same rate. This means that a bias of $1-\lambda^{-(d+2)}=\lambda^{-(d+1)}$ is optimal for ($d,d+1$) bit flipping, as stated in Case 2) of the theorem.  

We remark that the optimality of bit flipping for the ($2,4$) constraint is possible only because of the binary capacity equality $C(2,4)=C(1,2)$. The reason is as follows. We have seen that bit stuffing and bit flipping achieve capacity for ($d,\infty$) and ($d,d+1$) constraints. In both these cases, there is exactly one state in the FSTD that has two outgoing branches. This implies that a single DT can provide the required degree of freedom, and is sufficient to generate maxentropic sequences. With $d=1$, we can hence generate maxentropic ($1,2$) sequences using either bit stuffing or bit flipping. We can transform a maxentropic ($1,2$) sequence to a maxentropic ($2,4$) sequence using the following two operations
\begin{itemize}
\item Replace the sequence of phrases $0^110^11$ with the ($2,4$) phrase $0^31$
\item Replace the sequence of phrases $0^110^21$ with the ($2,4$) phrase $0^41$
\end{itemize}

\noindent This is equivalent to saying that since $C(2,4)=C(1,2)$, we have $\lambda_{2,4}=\lambda_{1,2}$ and hence the maxentropic $0^31$ and $0^41$ phrase probabilities can be written as, $\lambda_{2,4}^{-4}=\lambda_{1,2}^{-2}\lambda_{1,2}^{-2}$ and $\lambda_{2,4}^{-5}=\lambda_{1,2}^{-2}\lambda_{1,2}^{-3}$, respectively. $\lambda_{1,2}^{-2}\lambda_{1,2}^{-2}$ denotes the probability of concatenated ($1,2$) phrases $0^110^11$, and $\lambda_{1,2}^{-2}\lambda_{1,2}^{-3}$ is the probability of concatenated ($1,2$) phrases $0^110^21$. Note that there is no rate loss in the two operations. 

\smallskip

\noindent {\bf Case 3:} $2\le j\le k-d-1$

The above range of $j$ implies that we are dealing only with constraints ($d,k$) for which $k\ge d+3$. As in the previous two cases, we can derive the set of conditions from Table \ref{list}.
\begin{eqnarray}
\label{deux3}& & p=\lambda^{-1}\\
\label{trois3}& & 1-p=\lambda^{-(d+1)}\\
\label{quatre3}& & p^{k-d}=\lambda^{-(k-j+1)}
\end{eqnarray}

\noindent Once again, the above three conditions cannot be simultaneously satisfied unless $k=\infty$. Hence, we conclude that SS($j$), $2\le j\le k-d-1$ cannot achieve capacity for any ($d,k$).

\smallskip

\noindent {\bf Case 4:} $j=k-d$ and $j\ge 2$

It was shown in Lemma \ref{letdol} that $j=k-d$ is optimal for $k=2d+1$. We will now show that ($d,2d+1$) are the only set of constraints for which SS($k-d$) is capacity achieving. From Table \ref{list} we note that the following conditions need to be satisfied for SS($k-d$) to be optimal for any given ($d,k$). Recall that $j\ge 2$ and therefore $k-d\ge 2$.
\begin{eqnarray}
\label{deux4}& & p=\lambda^{-1}\\
\label{trois4}& & 1-p=\lambda^{-(d+2)}\\
\label{quatre4}& & p^{k-d}=\lambda^{-(d+1)}
\end{eqnarray}

\noindent From (\ref{quatre4}) and (\ref{deux4}) above, we require that $k-d=d+1$ or $k=2d+1$. It turns out (see Lemma \ref{letdol}) that this value of $k$ satisfies condition (\ref{trois4}) by virtue of the binary capacity equality $C(d,2d+1)=C(d+1,\infty)$. 

The Theorem statement now follows from Cases 1 through 4 above. Note that in the process, we have also shown that for all constraints ($d,k$), $k\ne d+1$, $k\ne \infty$, $k\ne 2d+1$ and ($d,k$)$\ne$ ($2,4$), the maximum average rate of SS($j$), $\forall$ $0\le j\le k-d$ is strictly less than capacity.
\end{proof}

\medskip

\begin{theorem}
\label{iff}
Let $0\le d<k<\infty$. Then for $0<j\le k-d$, the average rate of SS($j$) is greater than the average rate of SS($j-1$) if and only if $p>\lambda_{j-1,\infty}^{-1}$.
\end{theorem}
\begin{proof}
Let us denote by $R_j(p,d,k)$ the average information rate of SS($j$) for a given constraint ($d,k$) and bias $p$. We then have
\beq
\label{basic}
R_j(p,d,k)=h(p)\frac{L_{in}}{L^j_{out}},
\eeq
\noindent where $L_{in}$ and $L^j_{out}$ represent the average word lengths at the input and output to the SS($j$) constrained encoder, respectively (see Fig. \ref{cenc}). It can be seen that $L_{in}$ does not depend on the sliding index and is identical for all $j$, $0\le j\le k-d$. Hence, for a given bias $p$, $L^j_{out}$ is the important factor in comparing the information rates of SS($j$) and SS($j-1$). It is given by
\beq
\label{loutonebefore}
L_{out}^j=\sum_{i=0}^{k-d} v_i^jl_i^j,
\eeq

\noindent where $l_i^j$ is the length of the codeword (or ($d,k$) constrained phrase) corresponding to the phrase probability $v_i^j$ listed in Table \ref{list}. For example, index $i=j-1$ has $v_{j-1}^j=p^{k-d-j}(1-p)$ and $l_{j-1}^j=k-d-j+1$. Now, consider the difference of average output word lengths $L^{j-1}_{out}-L^j_{out}$. This is computed from (\ref{loutonebefore}) to be
\beq
\label{unwanted}
L^{j-1}_{out}-L^j_{out} = p^{(k-d)}-p^{(k-d-j)}(1-p)
\eeq

\noindent From (\ref{basic}) and (\ref{unwanted}), we can derive the condition, $R_j(p,d,k)>R_{j-1}(p,d,k)$ if and only if $p^j+p>1$. The proof is now completed using the fact that the only real, positive root of $p^j+p=1$ is $\lambda_{j-1,\infty}^{-1}$. 
\end{proof}  

\medskip

As a consequence of Theorem \ref{iff}, we state that if the rate maximizing bias for SS($j-1$) is greater than $\lambda_{j-1,\infty}^{-1}$, then SS($j$) achieves a higher maximum information rate than SS($j-1$) for the given ($d,k$) constraint. 

\medskip

\begin{theorem}
\label{rateex}
The average information rate of SS($j$) is given by
\begin{displaymath}
R_j(p,d,k)=h(p)\frac{1-p^{k-d}}{1-p^{k-d}+(1-p)\left(p^{k-d-j}-jp^{k-d}+d\right)}
\end{displaymath}
\end{theorem}

\smallskip

\begin{proof}
We start with (\ref{basic}) wherein
\begin{displaymath}
R_j(p,d,k)=h(p)\frac{L_{in}}{L^j_{out}},
\end{displaymath}
\noindent and write out the expressions for $L_{in}$ and $L_{out}^j$. $L_{in}$ is the average message word length into the constrained encoder of Fig. \ref{cenc}. Since it is independent of the sliding index $j$, we set $j=0$ and compute $L_{in}$ from Table \ref{bitstuff}. It is given by
\beq
\label{linone}
L_{in}=\sum_{i=0}^{k-d} v_i^0l_i
\eeq

\noindent where $l_i$ is the length of the corresponding message word listed in Table \ref{bitstuff}. For example, index $i=k-d-1$ has $v_i^0=p(1-p)$ and $l_i=2$. Writing this out, we obtain

\begin{eqnarray}
L_{in} & = & \sum_{i=0}^{k-d} v_i^0l_i\\
\label{yahaan} & = & 1-p+2p(1-p)+3p^2(1-p)+\ldots+(k-d)p^{k-d-1}(1-p)+(k-d)p^{k-d}\\
\label{oneplusp} & = & 1+p+p^2+p^3+p^4+\ldots+p^{k-d-1}\\
\label{oneminusp} & = & \frac{1-p^{k-d}}{1-p}
\end{eqnarray}

\noindent where (\ref{oneplusp}) is a direct simplification of (\ref{yahaan}).

Similarly, we now write out the expression for $L_{out}^j$, the average codeword length at the output of the constrained encoder. Clearly, this is dependent on the sliding index $j$. We start with the expression in (\ref{loutonebefore}) and write out the individual terms.\begin{eqnarray}
L_{out}^j & = & \sum_{i=0}^{k-d} v_i^jl_i^j\\
\label{yahaan1} & = & (1-p)(d+1)+p(1-p)(d+2)+p^2(1-p)(d+3)+\ldots+p^{k-d-j-1}(1-p)(k-j)\\
\nonumber & & +p^{k-d}(k-j+1)+p^{k-d-j}(1-p)(k-j+2)+\ldots+p^{k-d-1}(1-p)(k+1)
\end{eqnarray}

\noindent Now let 
\begin{eqnarray}
\mathcal{S} & = & L_{out}^0\\
\label{onnu} & = & (1-p)(d+1)+p(1-p)(d+2)+\ldots+p^{k-d-1}(1-p)k+p^{k-d}(k+1)\\
\label{rendu} & = & d+1+p+p^2+p^3+\ldots+p^{k-d}\\
\label{moonu} & = & d+\frac{1-p^{k-d+1}}{1-p}
\end{eqnarray}

\noindent where (\ref{onnu}) follows from (\ref{yahaan1}) with $j=0$. Using (\ref{yahaan1}), (\ref{onnu}) and (\ref{moonu}), we get
\begin{eqnarray}
\label{onnu1}L_{out}^j & = & (1-p)(d+1)+p(1-p)(d+2)+p^2(1-p)(d+3)+\ldots+p^{k-d-j-1}(1-p)(k-j)\\
\nonumber & & +p^{k-d}(k-j+1)+p^{k-d-j}(1-p)(k-j+2)+\ldots+p^{k-d-1}(1-p)(k+1)\\
\label{rendu1} & = & \mathcal{S}-jp^{k-d}+p^{k-d-j}-p^{k-d}\\
\\
\label{moonu1} & = & d+\frac{1-p^{k-d+1}}{1-p}-jp^{k-d}+p^{k-d-j}-p^{k-d}\\
\label{naalu1} & = & \frac{1-p^{k-d}+(1-p)\left(p^{k-d-j}-jp^{k-d}+d\right)}{1-p}
\end{eqnarray}

\noindent Substituting (\ref{oneminusp}) and (\ref{naalu1}) into (\ref{basic}), we obtain the expression for information rate as
\beq
\label{finalexp}
R_j(p,d,k)=h(p)\frac{1-p^{k-d}}{1-p^{k-d}+(1-p)\left(p^{k-d-j}-jp^{k-d}+d\right)} 
\eeq
\end{proof}

\medskip

In Theorem \ref{rateex}, we obtained an expression for the average information rate of SS($j$) in terms of the bias $p$, sliding index $j$ and constraint parameters $d,k$. For a given constraint ($d,k$), we are now interested in determining the values of $p$ and $j$ that jointly maximize $R_j(p,d,k)$. However, the complexity of the rate expression in (\ref{finalexp}) makes further analysis difficult. For this reason, optimization for both $p$ and $j$ is done numerically. Rate improvements for some important constraints are summarized in Table \ref{egpc}.
\begin{table*}[ht!]
\caption{Simulation Results of Rate Improvements for some constraints}
\label{egpc}
\begin{center} 
\small 
\begin{tabular}{|c|c|c|c|c|c|} \hline
{\bf ($d,k$)} & {\bf Shannon capacity} & {\bf Maximum bit} & {\bf Maximum bit} & {\bf Maximum symbol} & {\bf Maximizing symbol}\\
 & {\bf \emph{C(d,k)}} & {\bf stuffing efficiency (\%)} & {\bf flipping efficiency (\%)} & {\bf sliding efficiency (\%)} & {\bf sliding index $j$}\\ \hline
\hline
($1,3$) & 0.5515 & 98.93 & 99.74 & 100 & 2\\ \hline
($1,7$)& 0.6793 & 99.42 & 99.79 & 99.79 & 1\\ \hline
($2,5$) & 0.4650 & 98.47 & 99.74 & 100 & 3\\ \hline
($2,10$) & 0.5418 & 99.39 & 99.70 & 99.87 & 2\\ \hline 
($3,6$) & 0.3746 & 98.23 & 99.57 & 99.89 & 2\\ \hline
($4,8$) & 0.3432 & 98.02 & 99.16 & 99.91 & 4\\ \hline
($5,9$) & 0.2979 & 97.82 & 98.89 & 99.77 & 3\\ \hline
\end{tabular}
\end{center}
\end{table*}

\section{Construction 2: Optimal Codes Using Interleaving}
\label{interl}

Thus far, in Sections \ref{bitstu} and \ref{symsli}, we have studied the bit stuffing, bit flipping algorithms and proposed the symbol sliding algorithm to generate ($d,k$) constrained sequences. All three of these constructions used a single DT to generate an appropriately  biased, $i.i.d$ bit stream, which was then encoded into constrained phrases. Recently, Wolf \cite{Aviran04} observed that with the use of multiple such DTs, optimal bit stuffing encoders could be constructed for all values of $d$ and $k$. The idea is to generate several distinct biased streams, one each for a state in the FSTD that has two outgoing branches (see Fig. \ref{maxentr}). Since the number of such states is $k-d$ for $k<\infty$, we need precisely that many DTs to construct optimal codes in this fashion. We will refer to this scheme as the \emph{multiple DT} construction. 

In this section, we show that certain classes of ($d,k$) constraints allow optimal encoding using fewer than $k-d$ DTs. This is derived from the factorization of characteristic ($d,k$) polynomials and can be implemented using interleaving. We first describe such a construction for ($d,d+2^m-1$) constraints, $1\le m<\infty$, and then generalize to other constraints. 

\subsection{Optimal ($d,d+2^m-1$) Codes, $1\le m<\infty$}
\label{optima}

In order to understand the idea behind our code construction, we first briefly review the relationship between factorization of characteristic ($d,k$) polynomials and interleaving. Recall that the characteristic polynomial of the ($d,k$) constraint, $k<\infty$,  is given by (\ref{charpoly})
\begin{displaymath}
H_{d,k}(z)=\sum_{j=d+1}^{k+1}z^{-j}
\end{displaymath}

\noindent From $\mathcal{Z}$-transforms, we know that $z^{-j}$ indicates a delay of $j$ time periods. For our use of $z^{-j}$, $j$ denotes phrase length in bits. Hence, the characteristic polynomial $H_{d,k}(z)$ is really indicative that a ($d,k$) sequence is the concatenation of independent phrases from the finite set $\mathcal{X}=\left\{0^d1,0^{d+1}1,\ldots,0^{k-1}1,0^k1\right\}$. As before, if $k=\infty$, then $\mathcal{X}=\left\{0,0^d1\right\}$. 

Factorization of $H_{d,k}(z)$ has the interpretation of interleaving phrases corresponding to the individual factors. For example, consider the characteristic polynomial of the ($1,4$) constraint, $H_{1,4}(z)=z^{-2}+z^{-3}+z^{-4}+z^{-5}$. This can be factored as $H_{1,4}(z)=\left(z^{-1}+z^{-2}\right)\left(z^{-1}+z^{-3}\right)=H_{1,\infty}(z)H_{2,\infty}(z)$. The term $\left(z^{-1}+z^{-2}\right)$ represents a source that independently produces phrases of length one or two bits (or a source that produces a $(1,\infty)$ constrained stream). Similarly, $\left(z^{-1}+z^{-3}\right)$ represents a source that independently produces phrases of length one or three bits (or a source that produces a $(2,\infty)$ constrained stream). Interleaving phrases from these two sources yields a sequence of independent, concatenated phrases of length two, three, four or five bits, which is in turn described by $z^{-2}+z^{-3}+z^{-4}+z^{-5}$, the characteristic ($1,4$) polynomial. This gives the equivalence between interleaving and factorization. Note that the interleaving is based on length of individual phrases and not their representations.

Now consider the characteristic polynomial of the ($d,d+2^m-1$) constraint, $H_{d,d+2^m-1}(z)=\sum_{j=d+1}^{d+2^m}z^{-j}$. This can be factored as 
\begin{eqnarray}
\label{ekko}
H_{d,d+2^m-1}(z) & = & \sum_{j=d+1}^{d+2^m}z^{-j}\\
\label{do} & = & z^{-(d+1)}\prod_{i=1}^m\left(1+z^{-2^{(i-1)}}\right)\\
\label{teen} & = & z^{-(d-m+1)}\prod_{i=1}^mH_{2^{i-1},\infty}(z)
\end{eqnarray}

\noindent (\ref{teen}) shows that $H_{d,d+2^m-1}(z)$ can be written as the product of $m$ characteristic ($d,k$) polynomials, each with $k=\infty$ and some $d>0$. The term $z^{-(d-m+1)}$ up front in (\ref{teen}) merely acts as additional ``delay'' (or phrase length). Our code constructions are applicable even when $(d-m+1)<0$ in (\ref{teen}).

It is known from a result of Bender and Wolf \cite{Bender93}, that the bit stuffing algorithm is optimal for all ($d,\infty$) constraints, $d>0$. Recall that bit stuffing uses just a single DT. Hence, optimal codes can be constructed for ($d,d+2^m-1$) constraints using exactly $m$ DTs, one each for factors $H_{2^{i-1},\infty}(z)$, $i=0,1,\ldots,m$ in (\ref{teen}), and then suitably interleaving and encoding the biased streams. This is in comparison to the $2^m-1$ DTs that would be needed with the \emph{multiple DT} construction. 
\begin{figure}[ht!]
  \centerline{\resizebox{4.5in}{!}{\input{il.pstex_t}}}
  \caption{\textbf{Block diagram of the ($d,d+2^m-1$) code construction by interleaving. {\rm $\lambda$} denotes the positive real root of $H_{d,d+2^m-1}(z)=1$.}}
  \label{il}
\end{figure}

We now describe how the interleaving and encoding can be performed so that the codes produced are optimal. The block diagram in Fig. \ref{il} outlines our construction. First, the input is split into $m$ distinct streams using a serial to parallel (S/P) converter. These $m$ streams then act as inputs to the $m$ DTs. As before, DT($x$) dentoes a distribution transformer that outputs a binary $i.i.d$ stream with bias $x$ ($Pr\left\{0\right\}=x$) in response to an unbiased, $i.i.d$, binary input stream. The bias of the $m$ DTs are chosen so as to generate maxentropic ($d,d+2^m-1$) constrained phrases out of the encoder. It is known from a result of Zehavi and Wolf \cite{Zehavi} that the maxentropic phrase probabilities are $\lambda^{-i}$ for a constrained phrase of length $i$. Hence, we work backwards to determine the bias of the $m$ DTs, which turn out to be $\frac{1}{1+\lambda^{-2^l}}$, $l=0,1,2,\ldots,m-1$.

The $m$ biased bit streams now act as inputs to the bit interleaver. The bit interleaver produces a binary sequence $\textbf{u}=\left(u_{1}u_{2}\ldots u_{m}\right)\in \left\{0,1\right\}^m$ by interleaving the $m$ biased streams one bit at a time, in the specified order ($u_{1}$ is the MSB and $u_{m}$ the LSB). Finally, the encoder maps the binary sequence $\textbf{u}$ of decimal value $j$ to the ($d,k$) constrained phrase $0^{d+j}1$ (string of $(d+j)$ zeros followed by a one), $j=0,1,2,\ldots,2^m-1$. Table \ref{encode} specifies such an encoder mapping for ($d,d+7$) constraints. The size of this table is $8$ in the example and $k-d+1=2^m$ in general. 
\begin{table}[ht!]
\caption{Encoder Mapping for the ($d,d+7$) constraint}
\label{encode}
\begin{center} 
\small 
\begin{tabular}{|c|c|} \hline
\textbf{Interleaved binary} & \textbf{Corresponding ($d,d+7$)}\\
\textbf{sequence \hspace{0.2cm} u}=($u_1 u_2 u_3$) & \textbf{constrained phrase}\\ \hline
000 & $0^d1$\\
001 & $0^{(d+1)}1$\\
010 & $0^{(d+2)}1$\\
011 & $0^{(d+3)}1$\\
100 & $0^{(d+4)}1$\\
101 & $0^{(d+5)}1$\\
110 & $0^{(d+6)}1$\\
111 & $0^{(d+7)}1$\\\hline
\end{tabular}
\end{center}
\end{table}

The construction described above requires $m$ DTs, one $m$-bit interleaver and one variable length encoder. Hence, the number of required DTs is $log_2(k-d+1)$, as opposed to $k-d$ with the multiple DT construction. Next, we prove the optimality of our code construction.

\medskip
\begin{theorem}
\label{thepro}
The encoding procedure outlined in Fig. \ref{il} constructs optimal ($d,d+2^m-1$) codes.
\end{theorem}

\noindent \begin{proof}
In our construction, the bias of the $m$ DTs were chosen so as to generate maxentropic ($d,d+2^m-1$) constrained phrases at the output. Hence, our codes are optimal by the maxentropic property. We provide a complete proof in the Appendix \ref{appendi}.
\end{proof}

\subsection{Other Constraints} 
\label{otherc}

We now extend the interleaving construction proposed in Section \ref{optima} to a wider class of ($d,k$) constraints. The idea is to derive appropriate factorizations for general characteristic ($d,k$) polynomials, $k<\infty$. As before, we start with the characteristic polynomial
\beq
\label{start}
H_{d,k}(z)=\sum_{j=d+1}^{k+1}z^{-j} 
\eeq

\noindent The number of terms in the summation in (\ref{start}) is equal to $k-d+1$. Let $k-d+1$ be factored into the product of primes as 
\beq
\label{prodprimes}
k-d+1=\prod_{i=1}^n P_i
\eeq

\noindent Now define $\eta_i=\prod_{j=1}^i P_i$, $i=1,2,\ldots,n$, with $\eta_0=1$. It follows that $H_{d,k}(z)$ can be factored as 
\beq
\label{factor}
H_{d,k}(z)=z^{-(d+1)}\prod_{i=1}^n F_{d,k}^i(z),
\eeq

\noindent where each $F_{d,k}^i(z)$, $i=1,2,\ldots,n$, is of the form
\beq
\label{Fform}
F_{d,k}^i(z)=1+z^{-\eta_{i-1}}+z^{-2\eta_{i-1}}+\ldots+z^{-(P_i-1)\eta_{i-1}}
\eeq

Each factor $F_{d,k}^i(z)$ has $P_i$ terms and can be realized using ($P_i-1$) DTs. Hence, the total number of DTs required is $\sum_{i=1}^{n} \left(P_i-1\right)$. As long as $k-d+1$ is not prime, and the number of factors $n$ is greater than one, this is strictly less than the $k-d$ DTs required in the multiple DT construction. As an example, we now describe in detail our construction for the ($0,11$) constraint.

The characteristic ($0,11$) polynomial can be factored as
\begin{eqnarray}
\label{eleven}
H_{0,11}(z) & = & \sum_{j=1}^{12}z^{-j}\\
\label{twelve} & = & z^{-1}\left(1+z^{-1}\right)\left(1+z^{-2}\right)\left(1+z^{-4}+z^{-8}\right)
\end{eqnarray}

\noindent Fig. \ref{general_ex} shows the code construction that uses 4 DTs, one 4-bit interleaver and one variable length encoder. The bias of the $4$ DTs are determined exactly as in Section \ref{optima} by working backwards from a maxentropic output. The DTs with bias $\frac{1}{1+\lambda^{-1}}$ and $\frac{1}{1+\lambda^{-2}}$ correspond to factors $\left(1+z^{-1}\right)$ and $\left(1+z^{-2}\right)$, respectively. The remaining two DTs with bias $\frac{1}{1+\lambda^{-4}}$ and $\frac{1}{1+\lambda^{-4}+\lambda^{-8}}$ both correspond to the factor $\left(1+z^{-4}+z^{-8}\right)$. 

The interleaver functionality is slightly more complex in this case. If $u_1=1$, the interleaver generates a binary sequence $\textbf{u}=\left(u_{1}u_{2}u_{3}u_{4}\right)$ by interleaving the $4$ biased streams one bit at a time in the specified order ($u_{1}$ is the MSB and $u_{4}$ the LSB). If $u_1=0$, the interleaver skips the second biased stream (shown in dotted lines) and outputs the binary sequence $\textbf{u}=\left(u_{1}u_{3}u_{4}\right)$. The encoder then maps the binary sequence $\textbf{u}$ to ($0,11$) constrained phrases as specified in Table \ref{encodeone}. The size of this table is $12$ for this example and $k-d+1$ in general. 
\begin{figure}[ht!]
  \centerline{\resizebox{4in}{!}{\input{general_ex.pstex_t}}}
  \caption{\textbf{Block diagram of the ($0,11$) code construction by interleaving. {\rm $\lambda$} denotes the positive real root of $H_{0,11}(z)=1$.}}
  \label{general_ex}
\end{figure}

\begin{table}[ht!]
\caption{Encoder mapping for the ($0,11$) constraint}
\label{encodeone}
\begin{center} 
\small 
\begin{tabular}{|c|c|} \hline
\textbf{Interleaved binary} & \textbf{Corresponding ($0,11$)}\\
\textbf{sequence \hspace{0.2cm} u} & \textbf{constrained phrase}\\ \hline
 000 & $1$\\
 001 & $01$\\
 010 & $0^{2}1$\\
 011 & $0^{3}1$\\
1000 & $0^{4}1$\\
1001 & $0^{5}1$\\
1010 & $0^{6}1$\\
1011 & $0^{7}1$\\
1100 & $0^{8}1$\\
1101 & $0^{9}1$\\
1110 & $0^{10}1$\\
1111 & $0^{11}1$\\\hline
\end{tabular}
\end{center}
\end{table}

The code construction described above requires $4$ DTs, as opposed to $11$ DTs required with the multiple DT construction. The proof of optimality of the code construction in Fig. \ref{general_ex} proceeds similarly to that of Theorem \ref{thepro} and is hence omitted in the interest of space. 

\section{conclusion}
\label{conclu}

We introduced two new code constructions for the ($d,k$) constraint. First, we proposed the symbol sliding algorithm, which improves bit stuffing and bit flipping rates, and additionally achieves capacity for ($d,2d+1$) constraints. The main idea behind symbol sliding is to generate constrained phrases with probabilities that closely match that of the maxentropic sequence. We showed that this can be done by switching phrase probabilities from the bit stuffing algorithm. Furthermore, symbol sliding requires just one distribution transformer (DT), thus maintaining the simplicity of bit stuffing and bit flipping. 

Our second construction was inspired by a recent generalization of bit stuffing proposed by Wolf \cite{Aviran04}, where $k-d$ biased bit streams are used to construct optimal ($d,k$) sequences for all $k<\infty$. Here, we observed that the factorization of certain characteristic ($d,k$) polynomials could be used to construct optimal codes with fewer than $k-d$ DTs. This scheme was implemented using interleaving. In particular, we showed that optimal ($d,d+2^m-1$) codes, $1\le m<\infty$, could be constructed using just $m$ DTs.

We note that the optimality of the two constructions proposed in this work have different origins, eventhough their implementations are linked through the bit stuffing algorithm. The optimality of symbol sliding for ($d,2d+1$) constraints is possible only because of the binary capacity equality $C(d,2d+1)=C(d+1,\infty)$, and the fact that bit stuffing with a single DT achieves ($d,\infty$) capacity. With our second construction based on interleaving, the proof of optimality lies in the factorization of characteristic ($d,k$) polynomials. Hence, with the two different origins of optimality, we believe that further improvements might be possible with a combination of symbol sliding and interleaving.

\section*{acknowledgement}

The authors wish to thank Dr. Erozan M. Kurtas for his help, support and insight into this problem, and Sharon Aviran for providing a copy of reference \cite{Aviran04}.

\begin{small}
\bibliographystyle{IEEEbib11}

\end{small}

\begin{appendix}
\label{appendi}

\emph{Proof of Theorem \ref{thepro}}: We will show that the average information rate of the code construction in Fig. \ref{il} equals the capacity of the ($d,d+2^m-1$) constraint. The average information rate is given by
\beq
R(d,d+2^m-1)=\sum_{i=0}^{m-1} \frac{h\left(\frac{1}{1+\lambda^{-2^i}}\right)}{L_{out}}
\eeq

\noindent where $L_{out}=\sum_{j=1}^{2^m} (d+j)\lambda^{-(d+j)}$ is the average phrase length at the output of the encoder. The capacity of the ($d,d+2^m-1$) constraint can be expressed as
\beq
C(d,d+2^m-1)=log_2\lambda=\sum_{j=1}^{2^m} \frac{\lambda^{-(d+j)}log_2\left(\lambda^{(d+j)}\right)}{L_{out}}
\eeq

\noindent Hence, we need to show that $\sum_{i=0}^{m-1}h\left(\frac{1}{1+\lambda^{-2^i}}\right)=\sum_{j=1}^{2^m}\lambda^{-(d+j)}log_2\left(\lambda^{(d+j)}\right)$. We will start with the $R.H.S=\sum_{j=1}^{2^m}\lambda^{-(d+j)}log_2\left(\lambda^{(d+j)}\right)$ and show that it is same as the $L.H.S=\sum_{i=0}^{m-1}h\left(\frac{1}{1+\lambda^{-2^i}}\right)$. 
\begin{eqnarray}
\label{onearr}
\sum_{j=1}^{2^m}\lambda^{-(d+j)}log_2\left(\lambda^{(d+j)}\right) & = & \sum_{i=0}^{m-1}log_2\left(1+\lambda^{-2^i}\right) + log_2\lambda\frac{\sum_{j=1}^{2^m-1} j\lambda^{-j}}{\prod_{i=0}^{m-1}\left(1+\lambda^{-2^i}\right)}\\
\nonumber\\
\label{twoarr}
& = & \sum_{i=0}^{m-1}log_2\left(1+\lambda^{-2^i}\right) + log_2\lambda\sum_{i=0}^{m-1}\frac{2^i\lambda^{-2^i}}{1+\lambda^{-2^i}}\\
\nonumber\\
\label{threearr}
& = & \sum_{i=0}^{m-1}\frac{1}{1+\lambda^{-2^i}}log_2\left(1+\lambda^{-2^i}\right) + \sum_{i=0}^{m-1}\frac{\lambda^{-2^i}}{1+\lambda^{-2^i}}log_2\left(\frac{1+\lambda^{-2^i}}{\lambda^{-2^i}}\right)\\
\nonumber\\
& = & \sum_{i=0}^{m-1}h\left(\frac{1}{1+\lambda^{-2^i}}\right) \nonumber
\end{eqnarray}

\noindent where (\ref{onearr}) follows from the substitution $\lambda^{-(d+j)}=\frac{\lambda^{-(j-1)}}{\prod_{i=0}^{m-1}\left(1+\lambda^{-2^i}\right)}$, (\ref{twoarr}) is a result of dividing out the second term in (\ref{onearr}), and (\ref{threearr}) is a regrouping of the terms in (\ref{twoarr}).

\end{appendix}

\end{document}